\documentclass{iau}
\usepackage{amsmath}
\usepackage{graphicx}
\usepackage{multirow}
\usepackage{soul}
\usepackage{subcaption} 
\setulcolor{red}

\newcommand\prl{{Phys. Rev. Lett.}}%
\def\apj{{ApJ}}

\def\apjl{{ApJ Letters}} 
\def\apj{{ApJ}} 
   
\def\mnras{{MNRAS}}

\def\apjs{{ApJ Supplement}}
\newcommand\ssr{{Space~Sci.~Rev.}}%

\newcommand{\Fig}[1]{Figure~\ref{#1}}

\def\bl{Babcock--Leighton}

\begin{document}

\doival{ }

\aopheadtitle{Proceedings IAU Symposium No.365, 2023}
\editors{A. Getling \&  L. Kitchatinov, eds.}

\title{Solar cycle variability induced by stochastic
fluctuations of BMR properties and at different
amounts of dynamo supercriticality
}

\author{Pawan Kumar}
\affiliation{ Department of Physics, Indian Institute of Technology (Banaras Hindu University), Varanasi 221005, India\\ email:\email{pawan.kumar.rs.phy18@itbhu.ac.in}}


\begin{abstract}
Understanding the irregular variation of the solar cycle is crucial due to its significant impact on global climates and the heliosphere. Since the polar magnetic field determines the amplitude of the next solar cycle, variations in the polar field can lead to fluctuations in the solar cycle. We have explored the variability of the solar cycle at different levels of dynamo supercriticality. We observe that the variability depends on the dynamo operation regime, with the near-critical regime exhibiting more variability than the supercritical regime. 
Furthermore, we have explored the effects of the irregular BMR properties (emergence rate, latitude, tilt, and flux) on the polar field and the solar cycle. We find that they all produce considerable variation in the solar cycle; however, the variation due to the tilt scatter is the largest.

\end{abstract}

\begin{keywords}
Bipolar Magnetic Regions, Polar Field, Solar Cycle, Solar Dynamo
\end{keywords}

\maketitle

\section{Introduction}
The irregular variation of the solar magnetic cycle is the characteristic feature of the solar magnetic field with a periodicity of about 11 years. In addition to this, short-term variations, often referred to as the "seasons of the sun,"  are also observed in the cycle \citep{Rieger, EG16, Biswas23}. These variations manifest in phenomena such as double peaks ( Gnevyshev peaks) in the solar cycle, timing of polar field reversal, duration of polar field and solar cycle, active regions, extended periods of low and high solar activity, and total solar irradiance  \citep{KMB18, mord20, Mord22}. Therefore, understanding the variable nature of the solar magnetic cycle is essential, as it has direct and indirect impacts on human society. \citep{Petrovay20}.

The underlying dynamo process in the sun can explain the variability of the cycle \citep{Kar14a, Kar23}. It is believed that solar dynamo is $\alpha\Omega$ type, and the magnetic field grows when the dynamo number exceeds a critical value which is defined by,
\begin{equation}
 D = \frac{\alpha_0 \Delta \Omega R_\odot^3}{\eta_0^2},
\label{eq1}
\end{equation}
where $\alpha_0$ is the measure of the $\alpha$ effect, $\eta_0$ is the turbulent magnetic diffusivity,
$\Delta \Omega $ is angular velocity variation, and $R_\odot$ is the solar radius.
There is a huge amount of literature on the understanding of solar cycle variability based on the dynamo process \citep{Cha20, Kar23} and, which can be broadly classified as 1) nonlinear dynamics and deterministic chaos 2) stochastic forcing of the dynamo.
In the \bl\ dynamo scenario,
the nonlineary primarily comes from the flux loss due to magnetic buoyancy \citep{BKC22}, latitude quenching \citep{Kar20}, and tilt quenching \citep{JK2020}. The variability in the magnetic cycle comes from the fluctuating part of the \bl\ dynamo, i.e., from the \bl\ mechanism of polar field generation, which involves randomness. This extensive randomness arises in the \bl\ mechanism from the irregular properties of Bipolar magnetic Regions (BMRs), mainly from tilt scatter \citep{OK13, JK2020}.
The other sources of irregularity in the solar dynamo of observed BMR properties are time delay in BMR emergence, variation in flux, and BMR emergence latitude. The time delay of successive BMR emergence depends on the strength of the toroidal magnetic field in the convection zone \citep{JPL10} and may cause solar cycle variation. Observations suggest that the mean latitude of BMR emergence on the solar surface is not uniform from cycle to cycle and depends on the strength of the cycle \citep{MKB17}, which can lead to solar cycle variability. The flux of BMRs is also not constant and varies irregularly throughout the cycle in the range $10^{21}$ to $10^{23}$ Mx \citep{Anu23} and can produce variation in the solar cycle.

\section{Model and Method} \label{sec:mod}
To see the effect of irregular properties of BMR and supercriticality on the solar cycle variation, we used 2D and 3D dynamo models. We used the Surface Flux Transport And Babcock-Leighton (STABLE) dynamo model to incorporate the observed irregular properties of BMR in the model. For a detailed description of the model, see \citep{KM17}. The STABLE dynamo model can be used as the Surface Flux Transport (SFT) model and the dynamo model. To use it as the SFT, we feed the synthetic BMR data into STABLE and see the evolution of the polar field. We generated the synthetic BMR through synthetic BMR generation code following the \citet{J18}; also see \citep{Kumar23}. The BMR produced by the STABLE model and synthetic BMR generation code follows the following properties. 1) The time delay of BMR emergence roughly follows the lognormal distribution. 2) The flux variation is also lognormally distributed.
However, to see the solar cycle variation at different level of supercriticality, we used the \bl\ type flux transport dynamo model \citep{Kar14a, Cha20}. For a detailed understanding of the models used in the study, see \citet{Kumar21b}.

\section{Results}
In this section, we will demonstrate the solar cycle and polar field variability by utilizing 2D and 3D dynamo simulations. Initially, we will explain dynamo supercriticality's influence on the solar cycle's variation. Subsequently, we will elucidate the effects of irregular BMR properties on the polar field and its consequential impact on cycle variability.
\subsection{Variability in the different operation regime of the dynamo}
\begin{figure}
\includegraphics[scale=0.27]{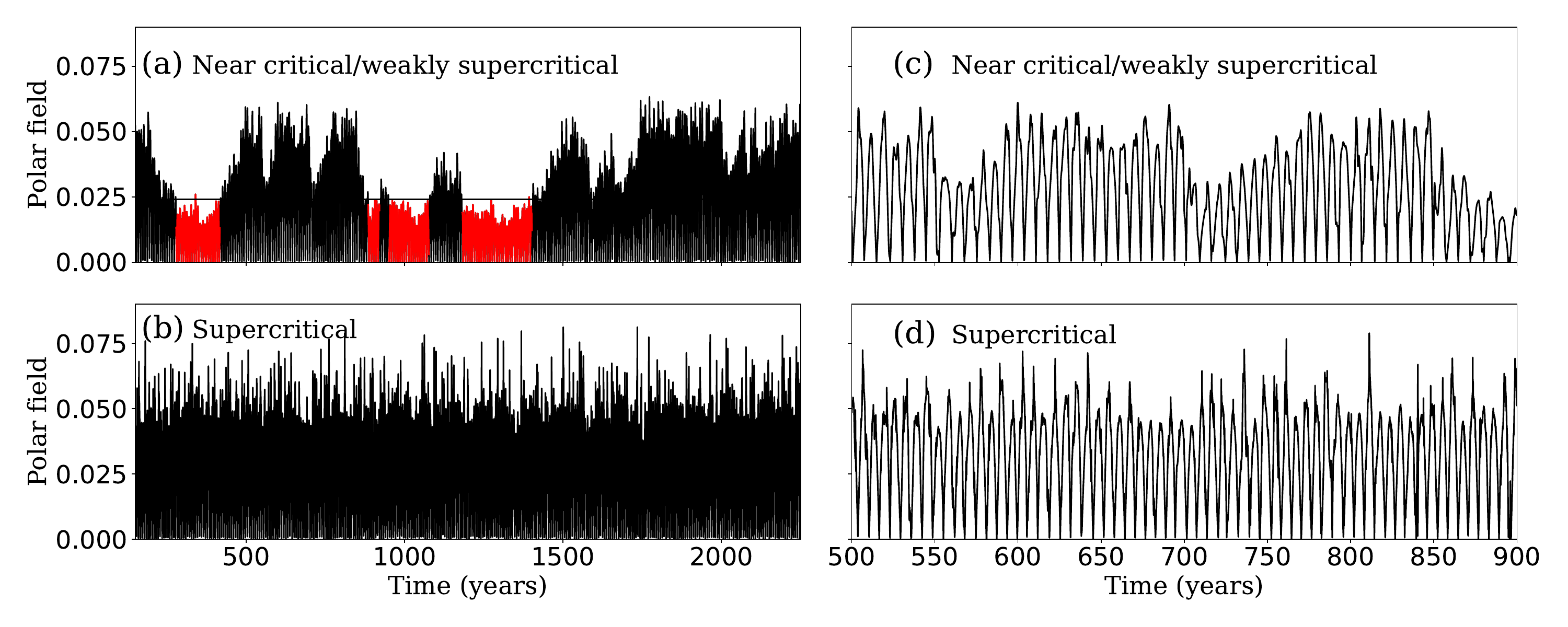}
\caption{The plot shows the variation in the solar cycle from a dynamo simulation which operates in the weakly supercritical region (a and c) and in the supercritical region (b and d). \textbf{c} and \textbf{d} show 400 years of data from the long Run \textbf{a} and \textbf{b}. The red color shows the extended episodes of grand minima in weakly supercritical regime. The plot are produced from the model used in \citet{Kumar21b}.} \label{fig:sup}
\end{figure}

We shall first examine the solar cycle variability across different levels of supercriticality in the dynamo, as illustrated in \Fig{fig:sup}. \Fig{fig:sup} (a and c) depict the variation in the solar cycle in the near-critical or weakly supercritical regime, while (b) and (d) illustrate the supercritical regime. The analysis of \Fig{fig:sup} reveals that when the dynamo operates in the weakly supercritical regime, it induces large variability in the cycle and extended episodes of grand minima \citep{C17, Kumar21b, V23}. 
On the other hand, when the dynamo operates near the supercritical regime, the variability in the cycle is notably reduced.
The cause of this is that when the dynamo operates near the critical regime, the inherent nonlinearity of the dynamo weakens. Consequently, even a slight change in the dynamo number ($D$) results in small magnetic field growth, as the impact is linear. Thus, after some time, when $D$ is increased due to fluctuations, the magnetic field experiences linear growth. Therefore, a large variability is observed in the magnetic field when the dynamo operates near the critical regime.
Conversely, in the supercritical regime, the dynamo's nonlinearity is strong. In this scenario, significant changes in $D$ do not translate into substantial growth in the magnetic field, as the nonlinearity suppresses such effects. Thus, in the supercritical regime, despite the large dynamo number, the variation in the magnetic field is observed less than the critical regime; also see \citep{KKB15, Vindya21, Kar23}.

\subsection{Variation in the polar field}
\begin{figure}[t]
\begin{minipage}[t]{0.55\textwidth}
 \centering
 \includegraphics[scale=0.32]{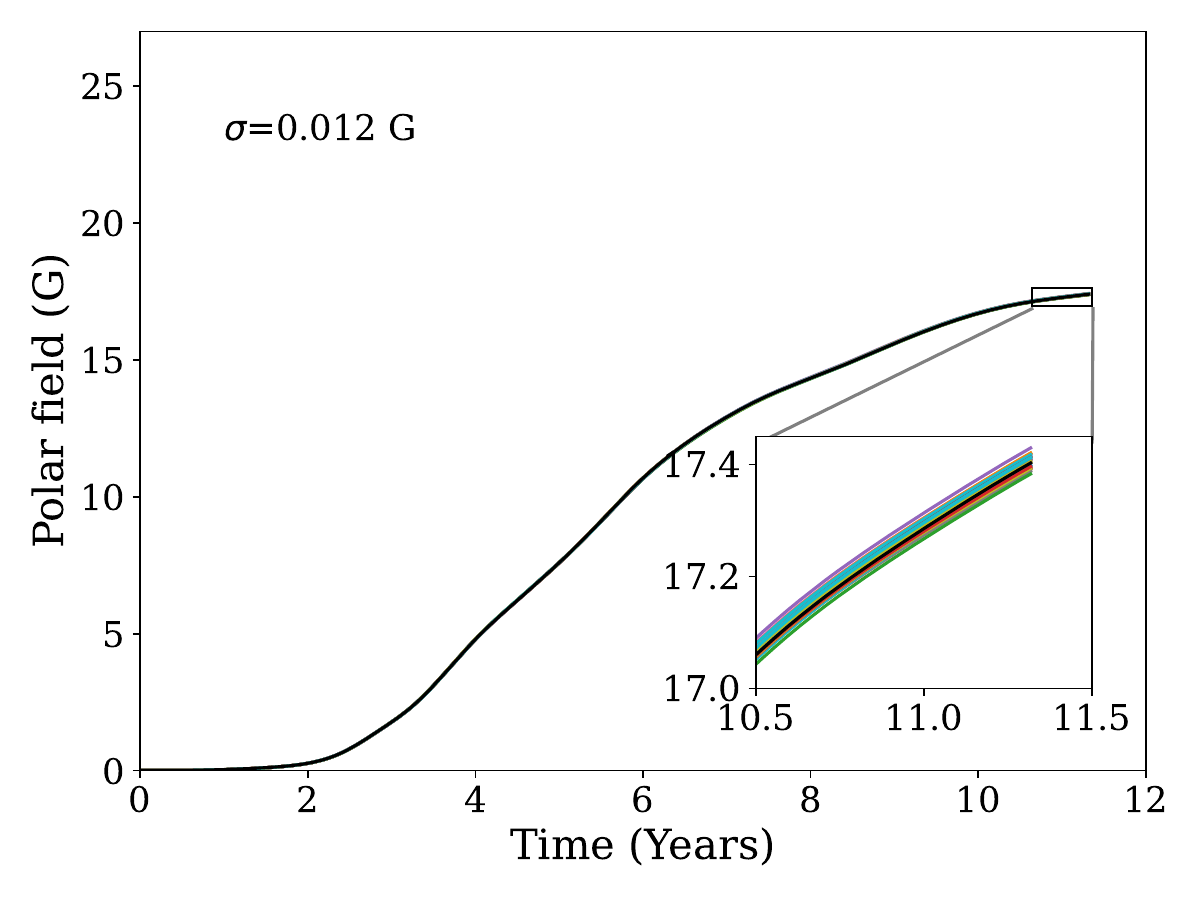}
 \end{minipage}%
 \begin{minipage}[t]{0.45\textwidth}
 \vspace{-4.5cm}
 \caption{ The plot illustrates the impact of time delays in successive BMR eruptions on the polar field. The black curve shows a reference polar field in the plot, with a zoomed-in view provided in the inset.}
 \label{fig:pol}
 \end{minipage}
\end{figure}

Now we shall discuss the variation in the polar field and in the solar cycle due to irregular properties of BMR.
Using the STABLE dynamo model as SFT, we see the variation in the polar due to different irregular properties of BMR. We see variations due to time delay in successive BMR emergence for different random realizations.
Due to time delay, we see a negligible variation in the polar field; see \Fig{fig:pol}. 
The reason for the negligible variation in the polar field due to time delays is that the maximum delay in BMR emergence is much smaller than the transport time of the flux towards the pole from lower latitudes. Consequently, the variation in the polar field due to time delay is smoothed out by the time required for the flux to reach the pole. As a result, the net change in the deposited flux in the polar region is negligible due to the smoothing effect of this transport time.

\begin{figure}[t]
\begin{minipage}[t]{0.8\textwidth}
 \centering
 \includegraphics[scale=0.28]{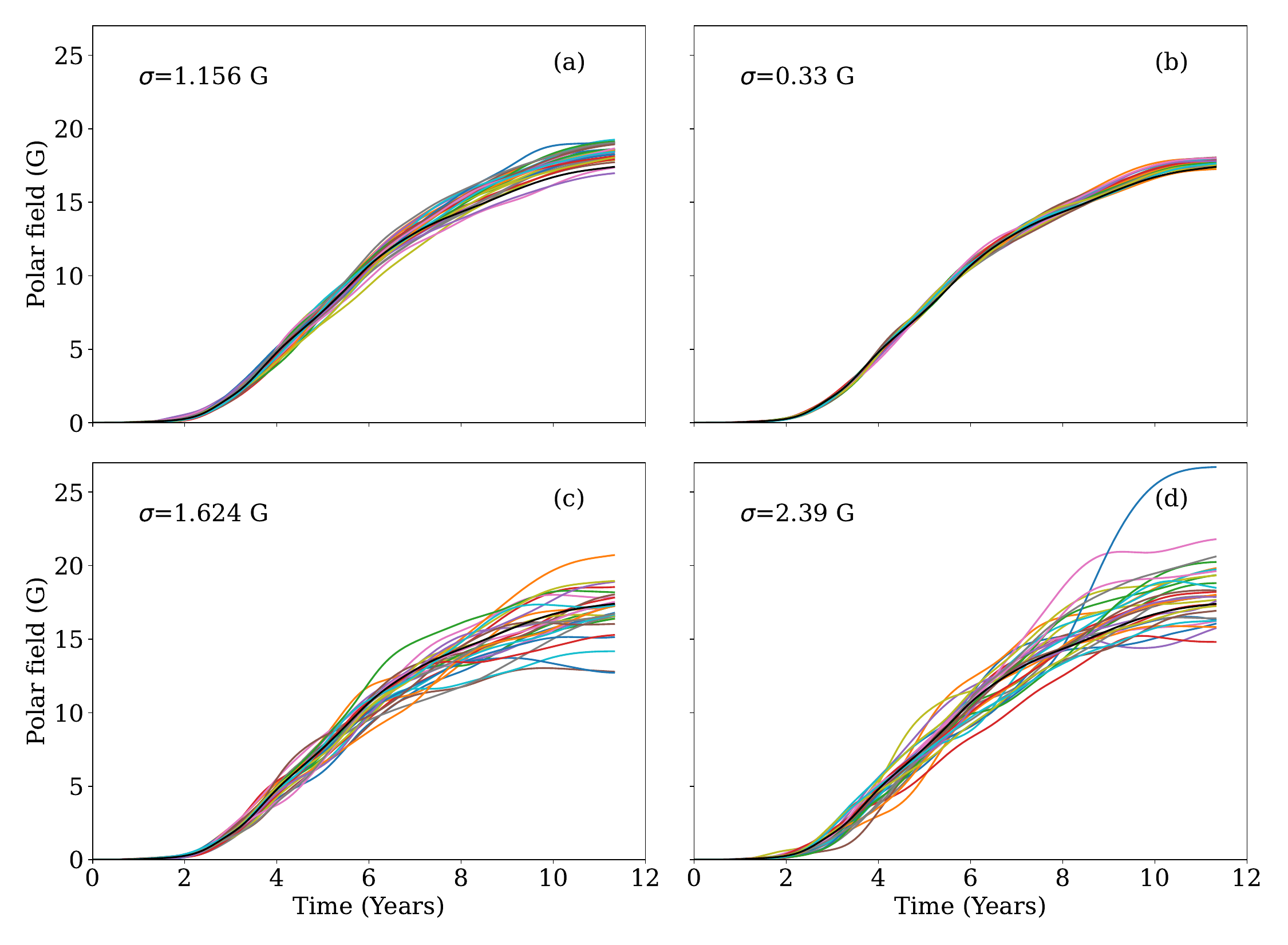}
 \end{minipage}%
 \begin{minipage}[t]{0.21\textwidth}
 \vspace{-7.5cm}
 \caption{Format of subplots are the same as in \Fig{fig:pol}, but shows the evolution of the polar field from simulations with (a) variation in the BMRs flux, (b) variation in BMRs emergence latitude, (c) scatter in BMR tilt, and (d) combined fluctuations.}
 \label{fig:pall}
 \end{minipage}
\end{figure}

\Fig{fig:pall} shows the variation in the polar field due to fluctuation in BMR flux, latitude, scatter in tilt, and due to overall fluctuation, (a), (b), (c), and (d), respectively. We see that the maximum variation in the polar field is produced due to scatter in the BMR tilt. However, fluctuation in mean BMR latitude and flux variation also produces a significant amount of variability in the polar field; see \Fig{fig:pall}.
The cause of the variation due to these parameters is the following in brief. The anticipated variation in the polar field due to latitude variation stems from the fact that when BMRs emerge at low latitudes, the cross-equatorial flux cancellation is notably more effective compared to those at higher latitudes \citep{KM18, Kar20}. Moreover, it is important to note that the flux contribution to polar field formation in a given hemisphere is directly proportional to the total active region flux for that hemisphere \citep{KO11}. Therefore, any changes in the flux can induce variations in the polar field.
In addition, the scatter in the tilt arises from wrongly (anti-Joy and anti-Hale) tilted BMRs \citep{SK2012}. These wrongly tilted BMRs engage in annihilation with the normal BMR field, leading to a significant variation in the polar field \citep{Nagy17, EB2023}.

\subsection{Variability in the solar cycle}
\begin{figure}
\centering
\includegraphics[scale=0.48]{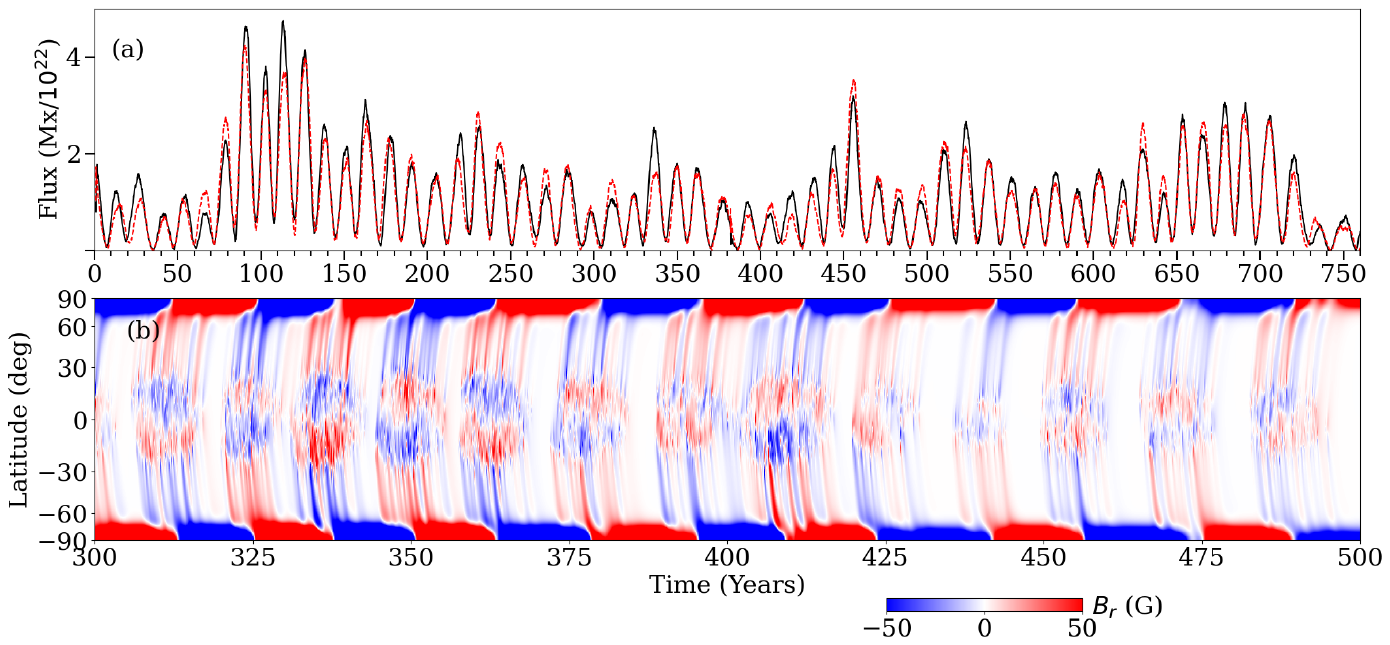}
\vspace{-2cm}
\caption{The plot shows the variation in the solar cycle from simulation, including all fluctuations (time delay, scatter in tilt, variation in latitude, and flux). Time delay is considered magnetic field dependent in this simulation.}
\label{fig:scy}
\end{figure}

The variability introduced in the solar cycle due to the irregular properties of BMRs is similar to that observed in the polar field. This result is evident because the polar magnetic field or its proxy strongly correlates with the next cycle's activity level \citep{Sch78, CCJ07, Kumar21, Kumar22, BKK23}. Similar to the polar field, the impact of time delay on solar cycle variation is found to be negligible. The maximum variability is produced due to the scatter in tilt, wherein BMRs with erroneously oriented Joy's and Hale's laws contribute significantly to fluctuations in the solar cycle.
Furthermore, the fluctuations in flux and emergence latitude also contribute substantially to the observed variation in the solar cycle. When all these fluctuations are considered collectively, the resulting variability in the solar cycle aligns consistently with the observed patterns, as depicted in \Fig{fig:scy}.

\section{Conclusion}
In our investigation, we utilized a 2D and 3D flux transport dynamo model to explore the variability in both the solar cycle and the polar field, considering the observed irregular properties of BMRs and different levels of supercriticality. The simulations revealed that, for the same level of fluctuation, if the dynamo operates near the critical regime, the resulting variation in the solar cycle is large. Conversely, when the dynamo operates near the supercritical regime, the observed variation in the solar cycle is comparatively small.
Furthermore, when we consider the irregular properties of BMR, the maximum variation produced in the polar field and the solar cycle is due to scatter in BMR tilt. The variation produced by the time delay is minimal; however, fluctuation in emergence latitude and the BMR flux produces significant variability in the polar field and the solar cycle. Moreover, tilt scatter in BMR tilt, flux variation, latitude variation, and fluctuation in time delay produce variability in the polar field and the solar cycle in the descending order, respectively.
These results also suggest that the irregular BMR properties sufficiently produce the variation in the polar field and the solar magnetic cycle, which is consistent with the observations.

\section{Acknowledgements}
The author thanks the International Astronomical Union (IAU) for providing the travel grant to attend the IAU Symposium 365 in Yerevan, Armenia. The author also acknowledges CSIR Symposia and Travel Grant Unit and IIT (BHU) for providing partial travel support to attend the IAU Symposium.

\bibliographystyle{iaulike}

\end{document}